\documentclass[10pt,a4paper,twoside]{article}
\usepackage[utf8]{inputenc}
\usepackage[T1]{fontenc}
\usepackage{lmodern}
\usepackage{booktabs}
\usepackage{graphicx}
\usepackage{nicefrac}
\usepackage[left=2cm,right=2cm,top=2cm,bottom=2cm]{geometry}
\author{P. Trybek $^{1*}$, M. Nowakowski $^{2}$, J. Salowka $^{3}$, J. Spiechowicz$^{4}$ and L. Machura $^{1}$}
\date{}
\title{Inter-pulse intervals of external anal sphincter surface EMG signals recorded from colorectal cancer patients}
\begin{document}
\maketitle
\noindent $^{1}$ \quad Division of Computational Physics and Electronics, Institute of Physics, 
Silesian Centre for Education and Interdisciplinary Research, University of Silesia in Katowice, Poland\\
$^{2}$ \quad Department of General Surgery and Multiorgan Trauma, Jagiellonian University Medical College, Krakow, Poland\\
$^{3}$ \quad Department of Surgery, Stanley Dudrick Memorial Hospital, Skawina, Poland\\
$^{4}$ \quad Department of Theoretical Physics, Institute of Physics, 
Silesian Centre for Education and Interdisciplinary Research, University of Silesia in Katowice, Poland\\
$^*$ paulina.trybek@smcebi.edu.pl

\section*{Abstract}
Information theory provides a spectrum of nonlinear methods capable of grasping an internal structure of a signal together with an insight into its complex nature. In this work, we discuss the usefulness of the selected entropy techniques for a description of the information carried by the surface electromyography signals during colorectal cancer treatment. The electrical activity of the  external anal sphincter can serve as a potential source of knowledge of the actual state of the patient who underwent a common surgery for rectal cancer in the form of anterior or lower anterior resection. The calculation of Sample entropy parameters has been extended to multiple time scales in terms of the Multiscale Sample Entropy. The specific values of the entropy measures and their dependence on the time scales were analyzed with regard to the time elapsed since the operation, the type of surgical treatment and also the different depths of the rectum canal. The Mann Whithney U test and Anova Friedman statistics indicate the statistically significant differences among all of stages of treatment and for all consecutive depths of rectum area for the estimated Sample Entropy. The further analysis at the multiple time scales signify the substantial differences among compared stages of treatment in the group of patients who underwent the lower anterior resection.

\noindent Keywords: surface electromyography, colorectal cancer, sample entropy, multiscale entropy

\section{Introduction} The comprehensive knowledge about the information hidden in the complex surface electromyographic (sEMG) 
signals of the external anal sphincter (EAS) could significantly contribute to the proper assessment 
of the activity of this specific muscle group in the context of patients after multimodal rectal cancer therapy. 
Colorectal cancer (CRC) is one of the most frequent cancers worldwide and nowadays represents a significant part of the 
major public health problems \cite{bray2012global}. 
Increasing morbidity and mortality rates indicate a rising global burden of CRC.  \cite{ferlay2015cancer}.
The latest predictions for 2030 estimate approximately 2.2 million new cases per year \cite{arnold2016global}. 
The standards of patient care require complex multimodal treatment composed of surgery, irradiation, and chemotherapy.
The medical protocol is strongly dependent on type, localization and the stage of CRC.
Especially the first two mentioned treatment modalities can have a significant  impact on the long term quality of life after the therapy 
due to their side effects. Those of special importance include stool and gas control and can range from minor gas leak to complete stool incontinence or evacuation difficulties. Frequency of those problems earned them even a separate name and are often referred to as (Low) Anterior Resection Syndrome (LARS) \cite{lin2015fecal}. 

It is also documented that the surgery, especially the level of anastomosis in conjunction with neoadjuvant radiotherapy, could increase the risk of postoperative 
complications associated with fecal incontinence \cite{bruheim2010late}. 
Despite many exhaustive reports about LARS \cite{juul2014low,ridolfi2016low} 
it is still not absolutely clear what kind of pathophysiologic mechanisms are mostly responsible for the postoperative dysfunction of the EAS muscle group. 
There are some suggestions that innervation injuries might have a relevant contribution to the multifractional LARS etiology \cite{rao1996anterior}.  

The distribution of innervation zones (IZ) shows a large discrepancy in the studied groups and relatively high level of individual 
patient asymmetry \cite{hinninghofen2003asymmetry,enck2004functional}. 
Thus, the difficulties in proper cognition of the main source of LARS are also dictated by the significant impact of intersubject variability. 
The other issues of great importance concern the different locations of anastomosis in the rectal area regarding the proximity of the 
sphincter muscles or the destructive effect of radiation.
All of these factors lead to the conclusion that the sphincter--sparing procedures require the thorough 
diagnostic tests of EAS neuromuscular system at every stage of the treatment process to be able to 
properly choose treatment regimens and assess risk factors. 
Among the applied techniques for monitoring the activity of EAS, considerable attention has been paid 
to the methods of electromyography (EMG).

Previous studies characterize the coaxial needle technique as an effective tool for investigating the 
neural control of EAS in the patient with defecation disorders \cite{podnar1999standardization}. 
However, due to some limitations of this method, mainly caused by its invasive character and 
technical difficulties related to a low repeatability of measurements (sampling error due to the placement of needle electrodes), 
the surface electromyography (sEMG) as a non--invasive equivalent has gained a wide range of application in this field 
\cite{enck2005external,mesin2009automatic,cescon2011geometry,cescon2014characterization,nowakowski2017sensitivity}. 
The exhaustive report of available techniques for acquiring the sEMG data from the external anal sphincter 
was presented by Merletti in \cite{merletti2016surface}. 
The continuous progress in the construction of measurement devices allows the gathering of new valuable information about 
the EAS motor units like the precise localization of the active innervation zones. Despite these experimental successes,
the literature still lacks the comprehensive theoretical characterization of raw sEMG in this specific clinical context.
sEMG signals always represent complex nature with low signal to noise ratio \cite{clancy2002sampling}. 
The ability to monitor the whole group of motor units from some distance entails, in turn, a negative cross-talk effect
due to the impact of the neighboring muscle activity. 
The tissue characteristics or the noise generated by external devices are among common factors which intensively influence 
the morphology of the signal wave.
 
To get a more profound insight into information hidden in sEMG, the use of proper analytical methods which can cope
with the complex character of the examined phenomena is required \cite{trybek2018multifractal}. 
The information theory with the special emphasis on 
the entropy-based techniques has become one of the very promising branches among the variety of algorithms used in 
biomedical signal processing \cite{beckers2001approximate,gao2015multiscale,zhang2013multiscale,trybek2018distribution}. 
Under normal healthy conditions, physiological systems are characterized by high dynamical complexity which is 
conditioned by their ability of quick adaptation to an incessantly changing environment. The loss
of such complexity is often related to the pathological state \cite{goldberger2002physiologic}.

The concept of entropy for a characterization of the measured data was first proposed in 1948 by 
Shannon in the form of the logarithmic dependence on a probability density function \cite{shannon1948mathematical}.
Further studies in this field resulted in the development of several forms of entropy measures,
from a notion of the Spectral Entropy, through the more advanced techniques like Approximate Entropy (AE) 
or its updated version Sample Entropy ($SampEn$) up to the Fuzzy Entropy presented by Chen et al. 
in 2007 \cite{chen2007characterization}.

A key limitation of these techniques is that they do not take into account multiple time scales. 
The biosignals often exhibit different behaviors depending on an actual scale.  
Nonlinearity, long memory or sensitivity to small disturbances are among the phenomena for which 
the description limited to a single time scale may not be sufficient. 
Although there exists a variety of entropy measures, the most widely used method in the 
context of a physiological signal's dynamics is Multiscale Entropy algorithm proposed by Costa 
et.al \cite{costa2002multiscale,costa2002multiscalebis,kang2009multiscale,humeau2015multiscale,faes2017efficient}.
In recent years several authors proposed an application of multiscale entropy and proved the method as a 
successive one for biomedical data analysis \cite{valencia2009refined,castiglioni2017multiscale}. One of the applications includes the description 
of the sEMG, i.e the activity of the urethral sphincter function \cite{wu2015multiscale} 
or a classification of the muscular disorders \cite{kaplanis2010multiscale}.
The aim of this work is to contrast the signals recorded at the different stages of rectal 
cancer treatment through the extensive analysis based on entropy parameters.
Both the specific values of the entropy measures and their dependencies on the time scales 
were analyzed due to factors such as the type of surgical treatment and the time of the 
recovery after the operation. Also the contraction and relaxation states at the different 
anatomical levels of the signal acquisition were considered separately.
 

\section{Methods}

\subsection{Sample Entropy}\label{SE}
The Sample Entropy ($SampEn$) represents the updated version of that developed by 
Pincus in 1991 Approximate Entropy ($ApEn$) \cite{pincus1991approximate}. There are several 
approaches for obtaining these entropy features. A brief description of the $SampEn$ 
algorithm used in this work is presented below. For more details, 
see \cite{richman2000physiological,costa2005multiscale}.

The calculation of $SampEn$ for the time series $\{x_i\}_{i=1}^N$ which consists of $N$ data points 
requires a prior determination of the two parameters: (i) the embedding dimension $m$ 
which characterizes the length of vectors to compare and (ii) the tolerance threshold $r$ 
referred to as a similarity criterion or the distance threshold for two template vectors. 
The latter is usually chosen from the range between $10\%$ and $20\%$ of the standard 
deviation $\sigma$ of the signal's amplitudes \cite{costa2003multiscale}.
In the following the values of $m=4$ and $r=0.2 \sigma$ have been used.

The procedure starts with the definition of a set of vectors $U_m(i)$ that represent $m$ 
consecutive values of series, starting with the $i-$th point
\begin{equation}
U_m(i)=\lbrace{x_i, x_{i+1}, \dots, x_{i+m-1}\rbrace}, \quad 1 \leq i \leq N-m+1 
\end{equation}
Next the Euclidean distance between the $U_m(i)$ and $U_m(j)$ is estimated as 
the absolute maximum difference between their scalar components:
\begin{equation}
d[U_m(i),U_m(j)]=\max_{k=0,\dots, m-1}(|x(i+k)-x(j+k)|)   
\end{equation}

In the next step the probability $C_i^m(r)$ that any $U_m(i)$ vector is close to $U_m(j)$ 
is determined. The $n_i^m(r)$ stands for a number of $U_m(j)$ vectors 
($1 \leq j \leq N-m$, $j \neq i$) that do not exceed the accepted tolerance 
threshold $r$ i.e. $d[U_m(i),U_m(j)] \leq r$. 
\begin{equation}
 C_i^m(r) = \frac{n_i^m(r)}{N-m} 
 \label{p1}
\end{equation}
This value is averaged over all possible pattern vectors $U_m(i)$ in order to estimate the probability 
$C^m(r)$ that any two vectors are within $r$ of each other
\begin{equation}
 C^m(r) = \frac{1}{N-m+1}\sum_{i=1}^{N-m+1}C_i^m(r) 
 \label{p2}
\end{equation}
%
%
Finally, the $SampEn$ is negative logarithm of the conditional 
probability that two sequences similar for $m$ points remain 
similar for the $m+1$ points.
\begin{equation}
 SampEn(m,r,N) = -\ln \left[ \frac{C^{m+1}(r)}{C^m(r)}\right ]    
\end{equation}
For the above calculations $j \neq i$, which means that self matches are not taken 
into account as in the case of earlier $ApEn$.

\subsection{Multiscale Entropy}
The estimation of Multiscale Entropy (MSE) consists of two main steps. 
The first part implements the coarse--graining 
procedure of resampling the series in order to explore 
different time scales of a signal \cite{semmlow2014biosignal}. 
The multiple coarse--grained time series are obtained 
by averaging the data points in each of the  non--overlapping windows with the increasing length. 
The procedure for the calculation of each of the coarse-grained series for the consecutive 
scale factors $\tau$ is given by
\begin{equation}
 y_j^{\tau} = \frac{1}{\tau} \sum_{i=(j-1)\tau+1}^{j\tau}x_i, \quad 1 \leq j \leq \frac{N}{\tau}    
\end{equation}
The second stage concerns the calculation of the sample entropy which was just presented in 
the previous paragraph (\ref{SE}). For each $y_j^{\tau}$ series the value of $SampEn$ is calculated and plotted as a function of $\tau$
resulting in the MSE curves.

\section{Material}\label{s2}
\subsection
{sEMG signal source}\label{surce}
\begin{figure}[htbp]
\centering
    \includegraphics[width=0.6\linewidth]{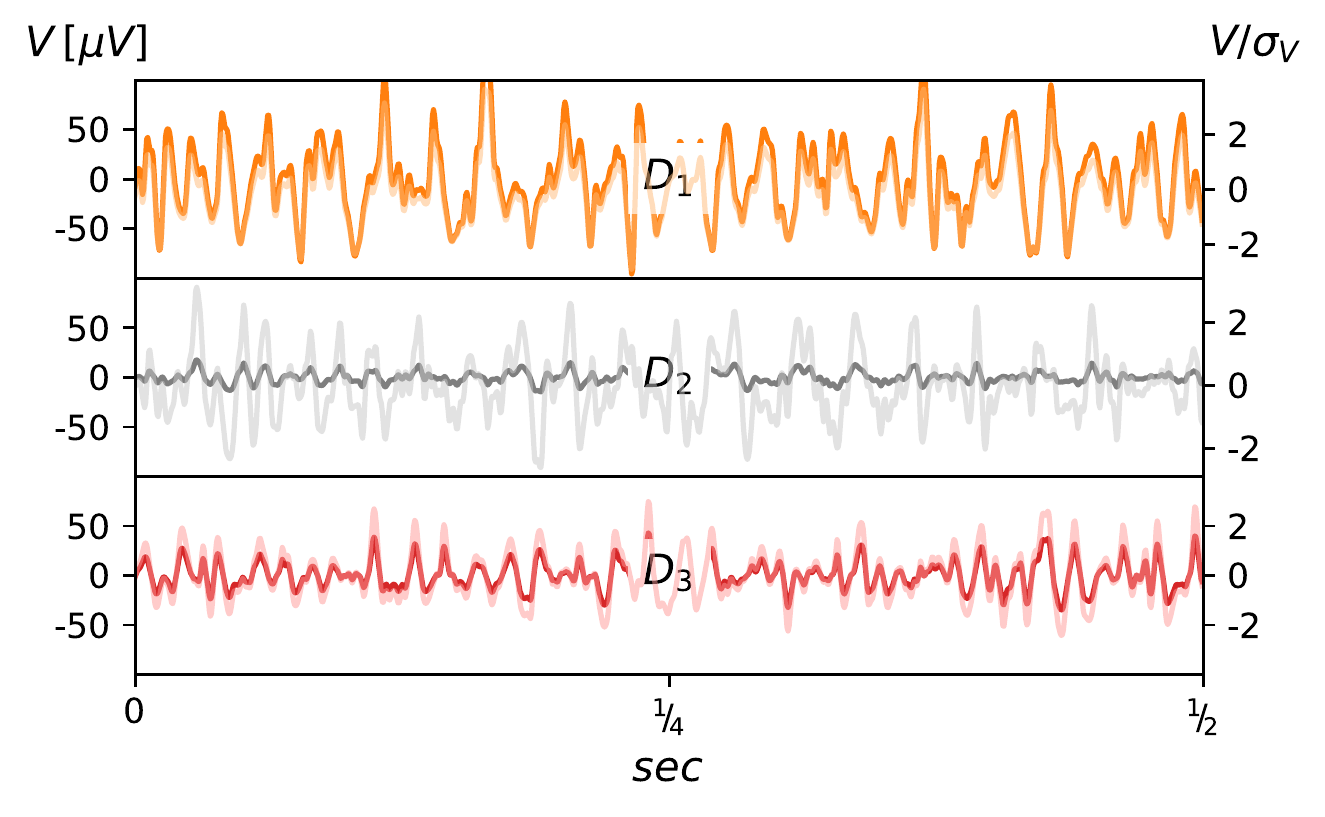}
    \caption{{The exemplary original (lhs scale, normal lines) and normalized (rhs scale, translucent lines) EMG data registered from the one of 16 channels at three different stages of treatment.}}
    \label{fig1}
\end{figure}

The examined time series were recorded at three stages of treatment, before the 
surgical procedure ($D_1$) and on two occasions in the postoperative period ($D_2$, $D_3$): 1 month after surgery $D_2$ and at 1 year $D_3$. {The exemplary raw and normalized EMG data are presented in} Fig.\ref{fig1}. {Normalization was performed with respect to the standard deviation, i.e. $\hat V = V / \sigma_V$.}
The data acquisition system consists of the anal probe developed at the Laboratory of Engineering 
of Neuromuscular System and Motor Rehabilitation, Politechnico di Torino in collaboration 
with the OT-Bioelettronica company. The signals were acquired from the 3 rings of 16 silver/silver oxide bar 
electrodes (1x10 mm) placed parallel to the long axis of the probe. 
Inter--ring distance was 8mm and that allowed for signal recording at approximate depths of 1, 3 and 5 cm from the anal verge.  The probe worked in conjunction with the standard 
PC over 12 bit NI DAQ MIO16 E-10 transducer (National Instruments, USA). The measurement protocol included 
respectively: 1 minute of relaxation, and three 10 sec long recordings at rest for each depth, 
1--minute relaxation, and then three 10 sec long recordings at maximum voluntary contraction (MVC) for each 
depth with additional 1--minute breaks in between. Each single 10 second-long measurement with the sampling 
frequency of 2048 Hz gave a series composed of 20480 data points. Low and high pass filters were used at 
10 and 500 Hz respectively. This resulted in typical 3dB bandwidth for the Analog-to-Digital Converter.

\subsection{Patients}\label{ss22}
{The study group included 20 subjects, 7 female, age range 46 to 71 
(average 57.14 $\pm$ 9.59  years) and 13 male, 
age range 48 to 85 (average 69.6 $\pm$ 10.04 years), diagnosed with a rectal cancer 
and qualified for surgery. 
All underwent open, transabdominal resection. Based on the distance of colo-rectal 
stapled anastomosis the study group of patients was divided according to the decision 
of the operating surgeon. For surgeons to make decision on the type of the procedure 
(AR vs LAR in this case) a localisation of the tumor is crucial. It is common to 
decide that tumors localised in the upper 3rd of the rectum require AR, 
those in middle portion LAR and those lower than that in some cases need LAR, 
ultra low LAR or abdomino-perineal resection.} 
The patients with anastomosis at, or below 6 cm from the dentate line were 
included to the Low Anterior Resection (LAR) group. 
Those with higher anastomosis were included to the Anterior Resection (AR) group. 
Indirectly, a level of anastomosis implies also the extent of mesorectal excision with all patients in LAR group undergoing Total Mesorectal Excision while in the case of AR group mesorectum was excised minimum 5 cm below the lower margin of tumor. 
For the detail information on the surgical landmarks of rectum see \cite{kenig2013definition}. 
{The group of patients is equally distributed with respect to the type of surgery: 10 subjects with AR (average years 62.4 $\pm$ 11.14) and 10 with LAR (average years 67.3 $\pm$ 9.97). The LAR group includes 8 males and 2 females, and except from one case, all the patients underwent the neoadjuvant radiotherapy (5x5 Gy for the total of 25Gy). The anterior resection group consists of 5 male and 5 female subject and none of those received neoadjuvant radiotherapy. In both groups TNM classification of patients was similar. 
In AR group there were 6 patients with T2 tumor and 4 patients with T3 tumor. In LAR group we had 5 patients with T2 and 5 with T3 tumors. In both cases resection is carried out along predefined planes and for the same localization T2 and T3 tumor should undergo the same resection. 
Regarding lymph node involvement 2 patients in AR group and 3 patients in LAR group where N1. All the others were N0. Regarding chemotherapy patients with positive lymph nodes received adjuvant chemotherapy. 
}


\section{Results}
\subsection{Choice of embedding dimension parameter}
\begin{figure}[htbp]
    \centering
    \includegraphics[width=0.75\linewidth]{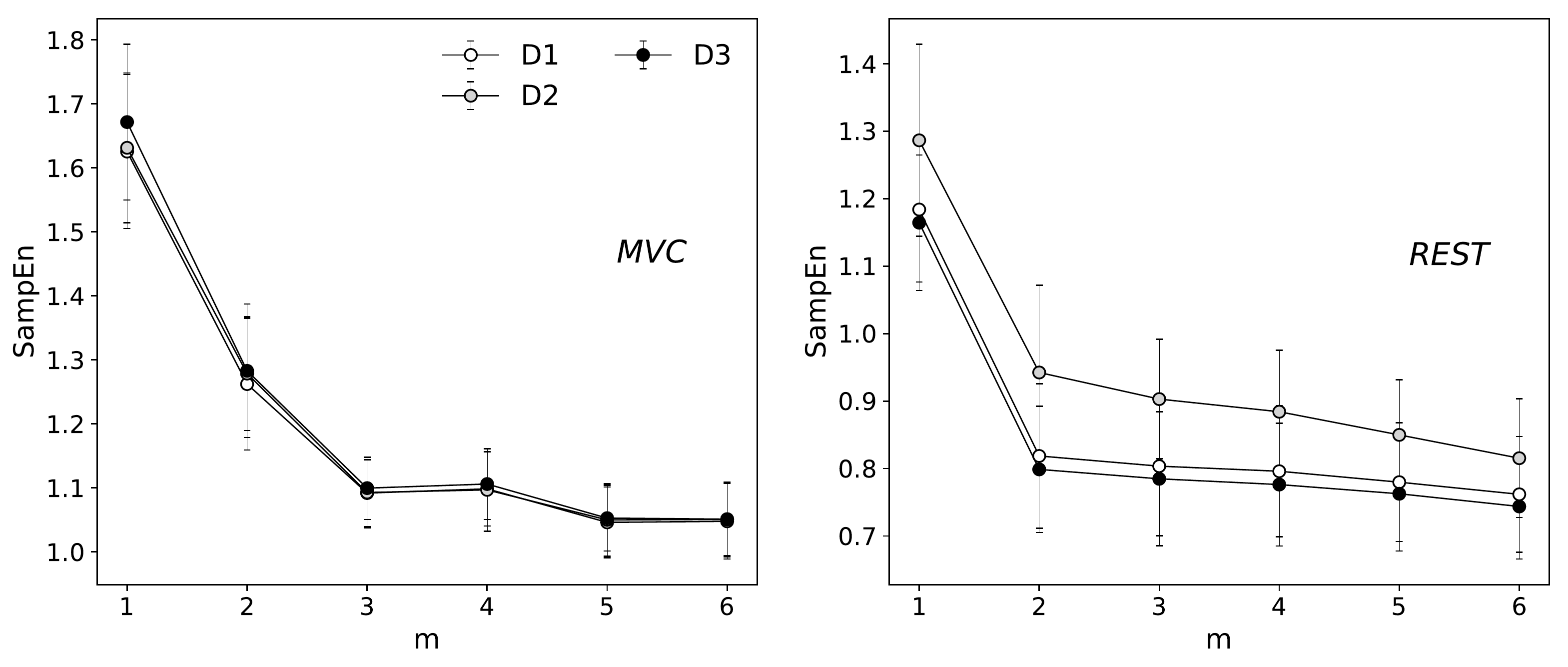}
    \caption{The Sample entropy calculated for the relaxation and maximum contraction state at the different embedding dimension setting. The left panel characterize the maximum contraction state; the right one is assigned to the relaxation.  The results are presented as an average of the 16-channels of selected state $D_1$.}
    \label{fig2}
\end{figure}

The standard protocol for the proper evaluation of motor units activity of the EAS muscle group recommends a minimum sampling frequency of about 2 kHz (\cite{merletti1999standards}, \cite{merletti2004electromyography}) and our data meets that restriction. In spite of that, the power spectra density estimation indicates the highest oscillations around 500 Hz (for details see \cite{trybek2018distribution}). 
{To eliminate the potential effect of overestimation of $SampEn$ through the comparing of the segments that consist of points with the same contribution to the signal, in other words, to avoid the situation that the four adjacent samples selected to form patterns we decided to choose $m=4$.} The effect of stabilization of $SampEn$ function along with the increase of embedding dimension is presented in Fig. (\ref{fig2}).

\subsection{Single scale entropy }

The results of $SampEn$ are presented 
in Fig.\ref{fig2}. There is a very clear 
division between the contrary states of EAS muscle tension. 

\begin{figure}[htbp]
    \centering
    \includegraphics[width=0.75\linewidth]{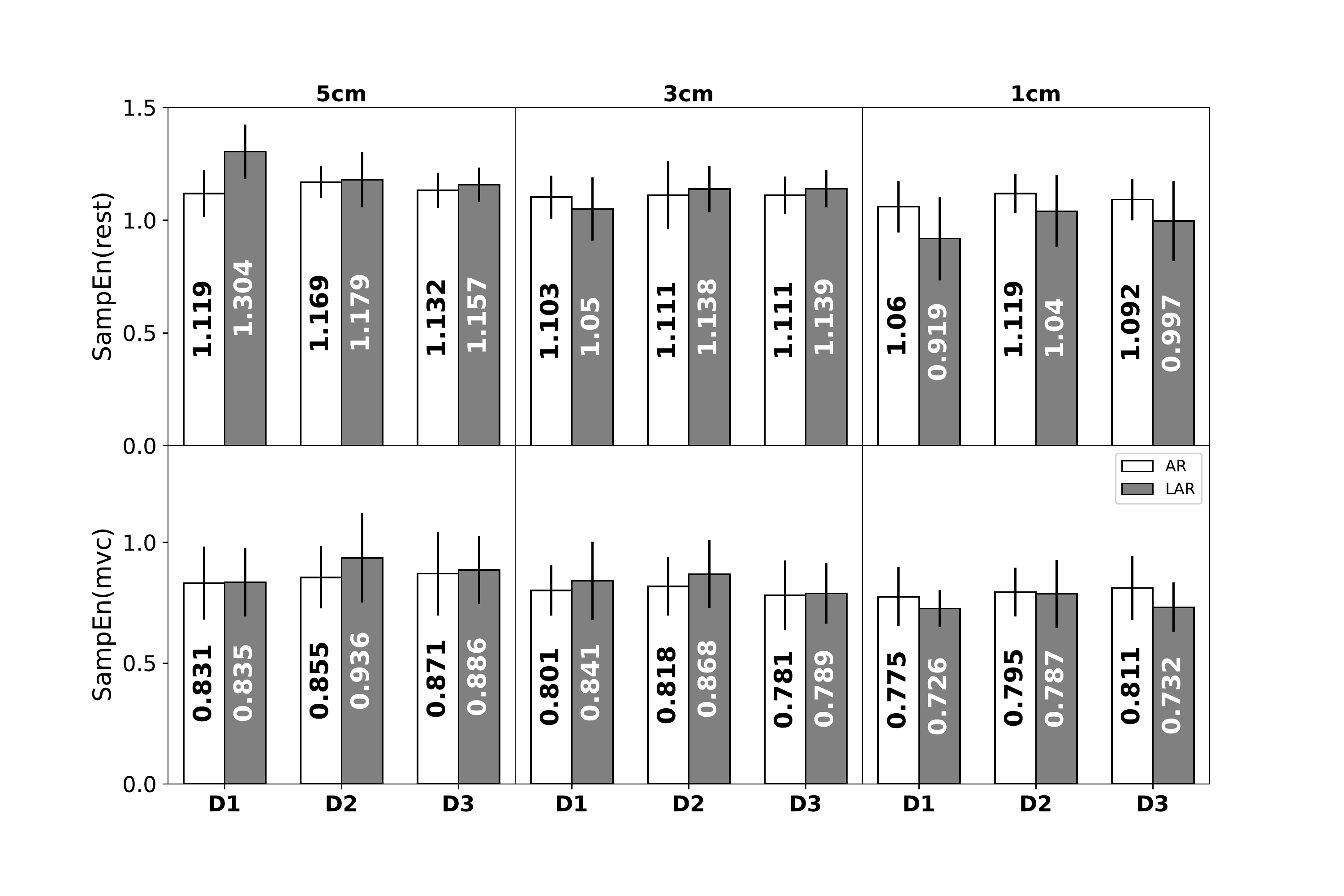}
    \caption{The average values of sample entropy calculated for patients with AR (white) and LAR(gray) at different stages of treatment($D_1-D_3$). The upper graphs represent respectively different depths of relaxation states. The lower charts are assigned to the maximum voluntary contraction.}
    \label{fig3}
\end{figure}

The significantly greater values of $SampEn$ for 
the relaxation state are partially justified by the concept of entropy as a measure of the diversity of the available 
states in the system. The reduction of these states is a consequence of the contraction phenomenon itself. During 
the propagation of the action potential within the functional motor units the specific direction of the process 
is dominated which automatically entails the decrease of the number of possible states that the system can choose. 
{The statistically significant differences calculated via Wilcoxon signed rank test at the selected significance level $\alpha=0.05$ were identified between the relaxation and maximum voluntary contraction in all the individual stages of treatment (D1-D3) and the respective depth of rectum canal (5cm-1cm). 
Interesting results concern the decrease in the mean values of sample entropy along with the rectum canal depth.  
Comparing the signals registered at 5cm and 1cm of depth in the case of contraction, the lower values of SampEn are assigned to the signals acquired in the immediate vicinity of a sphincter.
The most visible differences between AR and LAR group seem to characterize the relaxation state at 1cm of depth. At this specific level of the rectum, the AR group is characterized by the higher values of $SampEn$. The analogical tendency is visible for the maximum voluntary contraction.}


\subsection{Multiscale Entropy (MSE)}\label{r03}
\begin{figure}[htbp]
\centering
\includegraphics[width=0.9\linewidth]{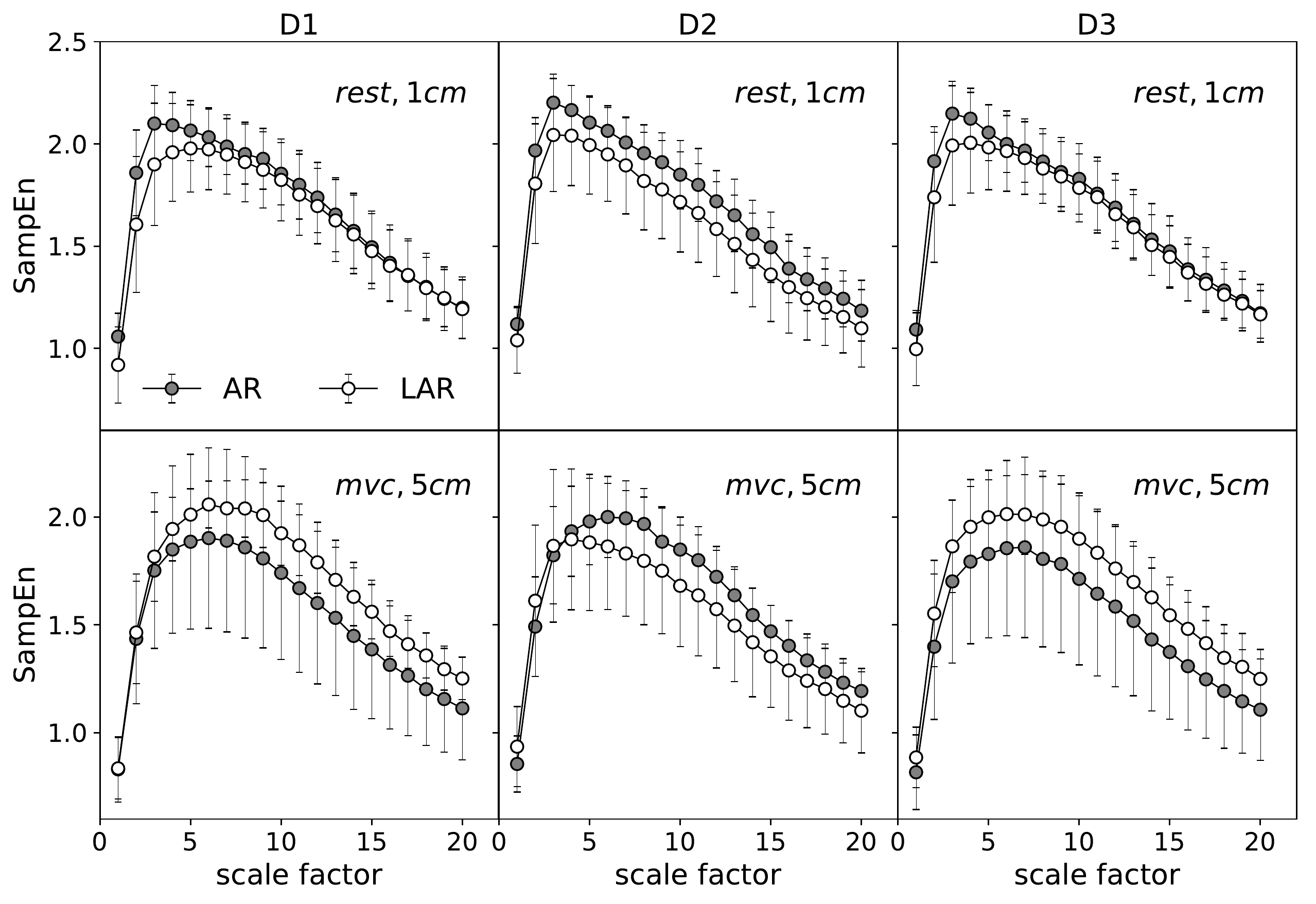}
\caption{The mean MSE entropy curve obtained for the selected cases: comparison of the group with AR and LAR for each of the treatment stages}
\label{fig4}
\end{figure}

The in--depth description of the examined data is given by $SampEn$ calculated over multiple timescales. 
Fig.\ref{fig4} gives an example of mean MSE curves for the selected representative stages. 
Each point presents an average of 160 values of $SampEn$ (16 signals per one subject) calculated for the 
respective coarse grained series at the consecutive scales $\tau \subset [1, 20]$. The differences of the MSE 
analysis between AR and LAR groups for the selected cases are illustrated.
The upper graphs represent the relaxation state at 1 cm of anal canal depth. 
The lower panels are assigned to the maximum voluntary contraction recorded at 5 cm. 
For both the relaxation and MVC $D_2$ significantly stands out from the other stages and there are no visible differences between the most distant stages of treatment $D_1$ and $D_3$.
The mean MSE curves of the state before surgery ($D_1$) and 1 year after operation ($D_3$) retain almost identical 
for the compared groups AR and LAR. Considering the different stages of muscle tension individually, the curves 
that illustrate the group with LAR are located respectively lower in the case of the relaxation state. The most visible 
differences identified at all ranges of scale occur one month after the surgery ($D_2$). 

\begin{figure}[htbp]
\centering
\includegraphics[width=0.9\linewidth]{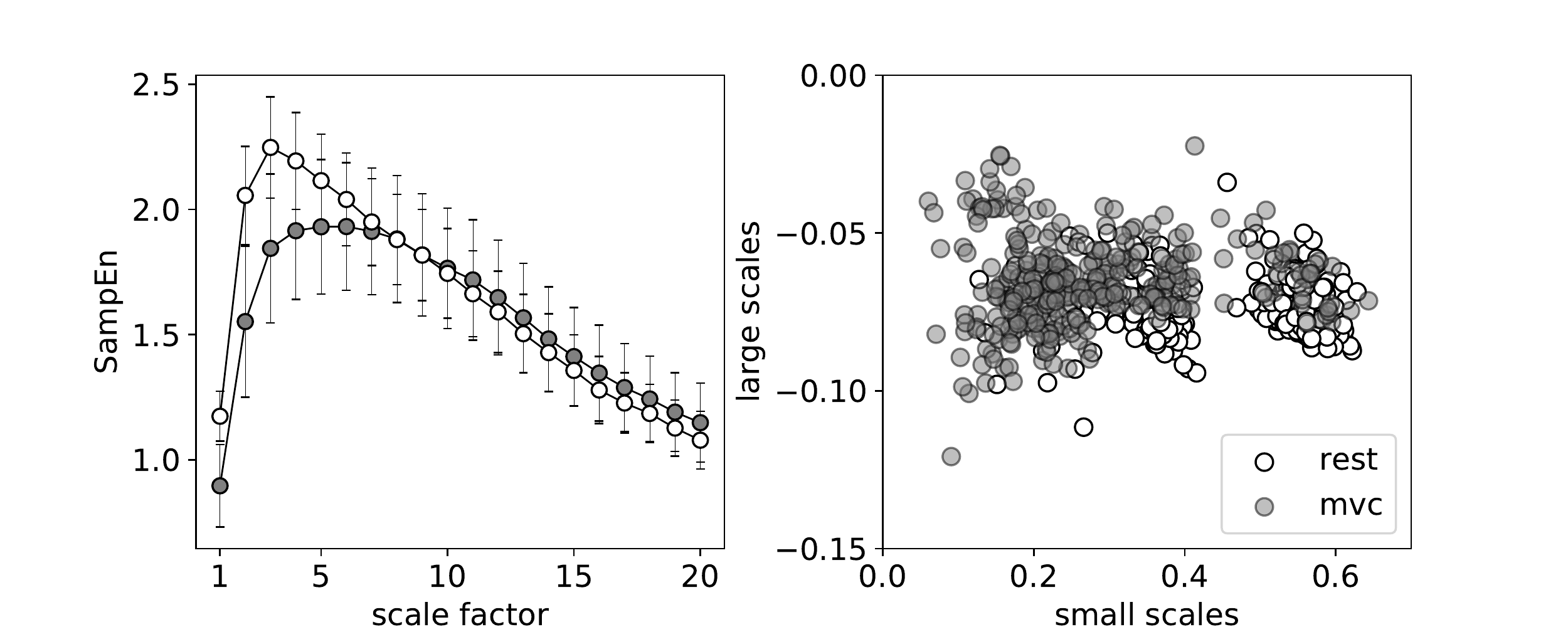}
\caption{The comparison of relaxation and MVC stage (an example calculated for both surgery group at 5 cm of depth one month after operation, l.h.s: MSE--curves, r.h.s: a point representation of MSE--curves.  }
\label{fig5}
\end{figure}

{Quite a different result is found for the MVC where the AR group is 
characterized by the reduced values of entropy in comparison to the LAR for $D_1$ and $D_3$ stages. Only the stage $D_2$ one month after surgery expresses higher values of $SampEn$ at the all scaling range.} 
In this case the group of patients with LAR is represented by the lower values of $SampEn$ for larger time scales. In general, the shape of curves assigned to the MVC and relaxation states have similar character. The rapid increase 
up to a certain maximum value at the relatively small scaling range and the monotonic decrease for the large scale factors. 
The observed differences are mainly manifested by the location of the maxima $\tau_{max}$. At relaxation state, the highest value of $SampEn$ is identified around the $\tau_{max}=3$ whereas for the contraction the maxima of MSE curves are shifted to the higher values of $\tau_{max} = 6$. The MVC--curves are also smoother around the maximum than in the case of relaxation. For better visualization of the differences between those contrary stages of muscle tension an example of MSE--curves of relaxation and MVC state are presented together on the left panel in Fig. \ref{fig5}. 

In addition the right plot presents a point representation of respective curves. Each point of the individual scatter plot 
is characterized by its respective coordinates. The abscissa represents the slope coefficients of the linear fit of the MSE curves for the small scales, i.e. the
ranges of scales between 1 and $\tau_{max}$. Accordingly, the ordinate is assigned to the slope coefficients of the linear fit for the large scales, i.e. $\tau > \tau_{max}$. Contrary to the relaxation in the case of  MVC the majority of points aggregate at the lower values of $\tau < \tau_{max}$. The slopes of the fit for the large scales are similar for both stages.


Considering the different stages of treatment the middle case -- $D_2$ appears to possess the highest variability among all presented cases. For this reason the comparison  of the stages before and after surgery individually for the LAR and AR groups needs to be addressed. In the following 
the $D_3$ stage will be omitted for the sake of the clarity of presentation.

\begin{figure}
\centering
\includegraphics[width=0.9\linewidth]{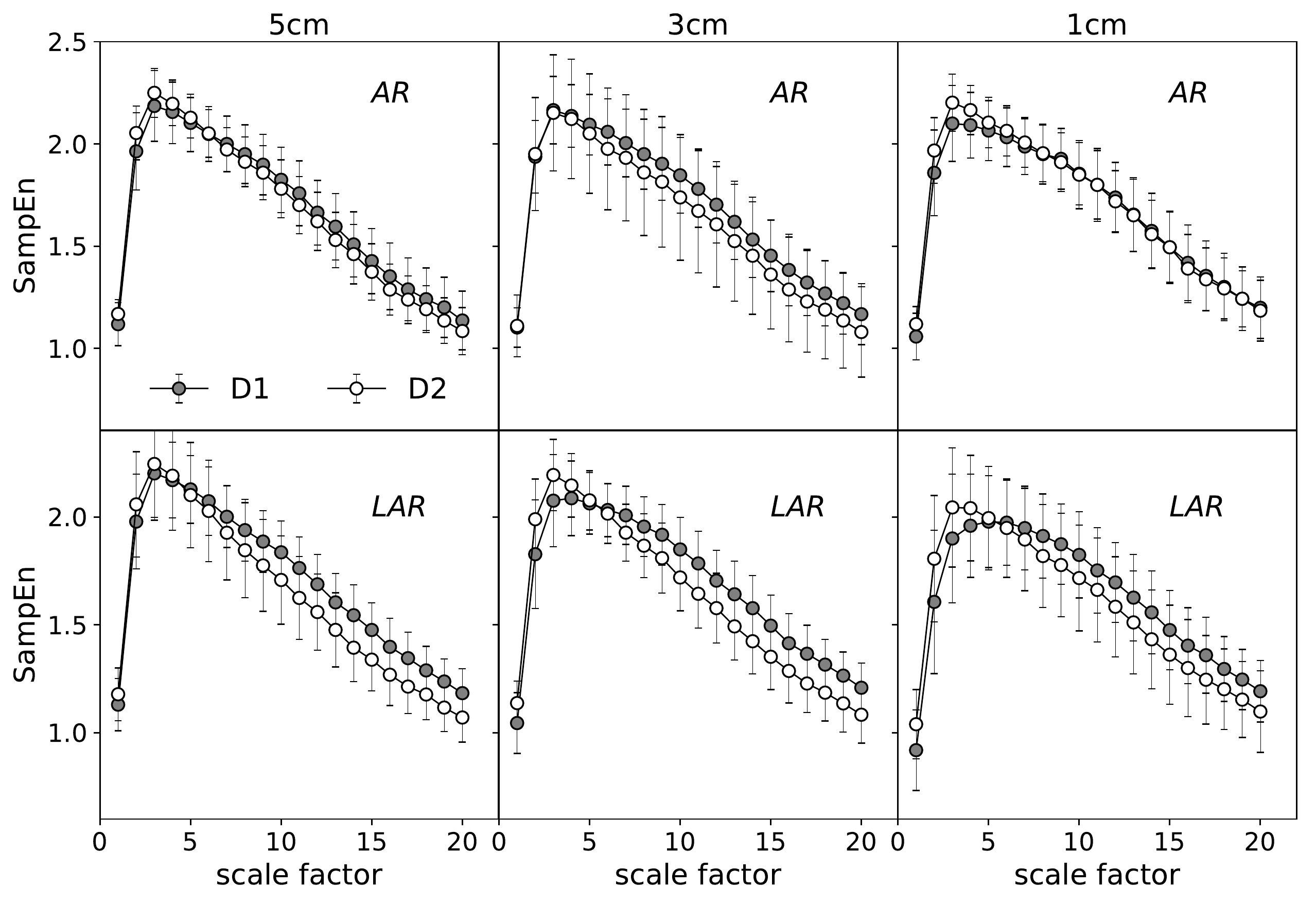}
\caption{The mean MSE entropy curves obtained for the relaxation state: comparison of stages $D1-D2$ in the group with AR and LAR}
\label{fig6}
\end{figure}

\begin{figure}
\centering
\includegraphics[width=0.9\linewidth]{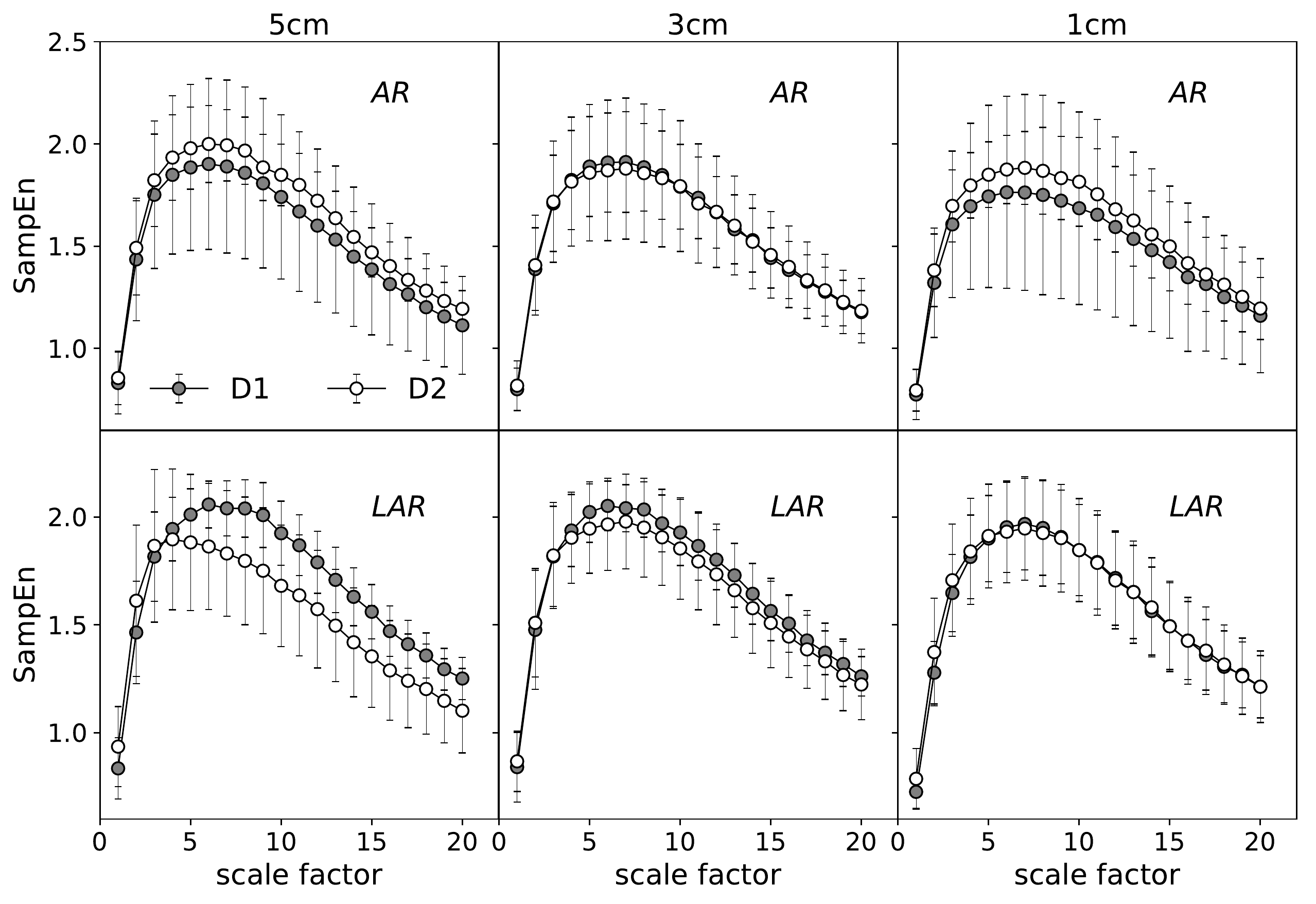}
\caption{The mean MSE entropy curves obtained for the MVC state: comparison of stages $D1-D2$ in the group with AR and LAR}
\label{fig7}
\end{figure}

Fig.\ref{fig6} present the results at the relaxation state in the form of mean MSE curves. The upper graphs characterize the AR group. The lower panels are their 
equivalent for the patients with LAR. The analogue set of results is given for the MVC (Fig.\ref{fig7}). 

A general comparison of the mean MSE curves characterizing the distinct states of EAS tension (Fig.\ref{fig6} vs Fig.\ref{fig7}) 
indicates the more visible differences between compared stages for the phase of relaxation. For this stage patients with LAR 
are characterized by the larger differences between $D_1$ and $D_2$ at all registered depths of rectum canal. The values of 
$SampEn$ are respectively lower for the $D_2$ over all considered range of scales. The  more pronounced differences seem to refer to the LAR group.  The contraction phase indicate visible differences between respective MSE curves at 5 cm of depth in the LAR group. In contrast to the AR group, $D2$ stage is characterized by the lower $SampEn$ values at the whole scaling range.

%


\section{Statistics}

The normality test of the entropy functions calculated via Shapiro--Wilk formula does not allow us to confirm the hypothesis 
about the normal distribution for the majority of analyzed cases. Thus, in order to characterize the differences between 
compared stages the non--parametric statistical tests were used.
The comparison between AR and LAR group are presented in table \ref{table1}. It consists of the results of the Mann-Whitney U 
test (the non--parametric equivalent of t--student statistics for the independent samples). The p--values, calculated at the 
selected significance level ($\alpha = 0.05$) are presented for each entropy measures. 
The statistically significant differences are featured in bold. The individual stages of treatment ($D1-D3$) together with 
the respective depths of rectum canal ($5cm$ - $1cm$) are taken into consideration. The table \ref{table1} set together results of single scale sample entropy and MSE considered as an average values of $SampEn$ at all scaling range.  The larger divergence between the AR and LAR groups is observed for the relaxation state at 1 and 5 cm of depth. For the multi--SampEn comparison, the MVC stage shows differentiation at the all compared states.

\begin{center}
\begin{table*}[h!]\centering
\caption{ The comparison of AR and LAR group: results of the non--parametric Mann–-Whitney U test calculated for entropy parameters.
The test was estimated for the MVC and the relaxation states at 1, 3 and 5 cm depth of the anal canal and consecutive stages of treatment: $D1$--$D3$.}
\begin{tabular}{@{}ccccccccc@{}}\toprule 
& \multicolumn{3}{c}{$SampEn$} & \phantom{abc}& \multicolumn{3}{c}{$MSE$ (all scales)} \\
rest & 5 & 3 & 1 && 5 & 3 & 1  \\ \midrule
$D_1$ & \textbf{0.050} & \textbf{0.001} & \textbf{0.000} && \textbf{0.027} & 0.579 & \textbf{0.000}\\
$D_2$ & \textbf{0.004} & 0.062 & \textbf{0.000} && \textbf{0.007} & 0.181 & \textbf{0.000}\\
$D_3$ & \textbf{0.000} & \textbf{0.001} & \textbf{0.000} && 0.901 & 0.245 & \textbf{0.000}  \\
MVC & 5 & 3 & 1 && 5 & 3 & 1  \\ \midrule
 $D_1$ & 0.602 & 0.228 & \textbf{0.000} && \textbf{0.000} & \textbf{0.000} & \textbf{0.000} \\
 $D_2$ & \textbf{0.000} & \textbf{0.004} & 0.464 && \textbf{0.000} & \textbf{0.000} & \textbf{0.000}\\
 $D_3$ & \textbf{0.000} & \textbf{0.001} & \textbf{0.000} && \textbf{0.000} & \textbf{0.000} & \textbf{0.000} \\ 
\bottomrule
\end{tabular}
\label{table1}
\end{table*}
\end{center}

Next, the statistical differences within the individual groups were also investigated. The results of Anova Friedman statistics, a widely known non–parametric analogue of the one-factor analysis of variance for the 
repeated measurements, indicates a statistically significant difference (p<0.05) between the consecutive stages of treatment at the individual depths of a rectum canal as well as the corresponding depths for the separate treatment periods. The respective differences were identified for both single scale $SampEn$ and MSE values. The single scale $SampEn$ results indicate statistically significant differences between all the comparing stages ($p<0.05$).  The full Friedman test characteristic of MSE values are presented in table \ref{table2} and \ref{table3} respectively.
{There are only two exceptions that do not allow us to reject the null hypothesis about the lack of statistically significant differences among the compared stages.} 
The state $D2$ at relaxation in the case of LAR group and $D1$ at MVC registered from the patient with AR are not diversified due to the depth of anal canal (see table \ref{table3}).



\begin{center}
\begin{table*}[h!]\centering
\caption{ The comparison of treatment stages at consecutive depth of rectum canal: full statistics of the non--parametric Friedman test calculated for $SampEn$ entropy at all scaling range.}
\begin{tabular}{@{}ccccccccc@{}}\toprule 
& \multicolumn{3}{c}{$AR$} & \phantom{abc}& \multicolumn{3}{c}{$LAR$} \\
rest & 5 & 3 & 1 && 5 & 3 & 1  \\ \midrule
 Anova $\chi^2$ & 57.07 & 106.64 & 86.14 && 413.45 & 384.08 & 43.51\\
Kendall coeff. & 0.009 & 0.017 & 0.014 &&   0.065 & 0.060 & 0.007\\
 p--value & \textbf{0.000} & \textbf{0.000} & \textbf{0.000} && \textbf{0.000} &\textbf{0.000} & \textbf{0.000}  \\
MVC & 5 & 3 & 1 && 5 & 3 & 1  \\ \midrule
 Anova $\chi^2$ & 110.17 & 55.77 & 261.24 && 517.23 & 302.40 & 55.61 \\
Kendall coeff. & 0.017 & 0.009 & 0.408 && 0.081 & 0.047 & 0.009\\
 p--value & \textbf{0.000} & \textbf{0.001} & \textbf{0.000} && \textbf{0.000} & \textbf{0.000} & \textbf{0.000} \\ 
\bottomrule
\end{tabular}
\label{table2}
\end{table*}
\end{center}

\begin{center}
\begin{table*}[h!]\centering
\caption{ The comparison of different depth of rectum canal at consecutive stages of treatment: full statistics of the non--parametric Friedman test calculated for $SampEn$ entropy at all scaling range.}
\begin{tabular}{@{}ccccccccc@{}}\toprule 
& \multicolumn{3}{c}{$AR$} & \phantom{abc}& \multicolumn{3}{c}{$LAR$} \\
rest & $D1$ & $D2$ & $D3$ && $D1$ & $D2$ & $D3$  \\ \midrule
 Anova $\chi^2$ & 24.05 & 173.95 & 8.612 && 110.88 & 2.974 & 28.67\\
Kendall coeff. & 0.004 & 0.027 & 0.001 &&   0.017 & 0.46E-3 & 0.004\\
 p--value & \textbf{0.000} & \textbf{0.000} & \textbf{0.013} && \textbf{0.000} & 0.226 & \textbf{0.000}  \\
MVC & 5 & 3 & 1 && 5 & 3 & 1  \\ \midrule
 Anova $\chi^2$ & 0.645 & 96.11 & 294.97 && 350.50 & 351.96 & 447.34 \\
Kendall coeff. & 0.1E-3 & 0.015 & 0.046 && 0.055 & 0.055 & 0.070\\
 p--value & 0.724 & \textbf{0.001} & \textbf{0.000} && \textbf{0.000} & \textbf{0.000} & \textbf{0.000} \\ 
\bottomrule
\end{tabular}
\label{table3}
\end{table*}
\end{center}

\section{Discussion}
This work presents an application of the selected entropy--based techniques to study the variability of information within 
sEMG signals at the different stages of the rectal cancer treatment. In order to distinguish the groups of patients due to 
the type of surgery as well as to compare of signals recorded at the various postoperative periods both, single and multiscale sample
entropy algorithms were implemented. 
The statistically significant differences identified among all of the compared stages of treatment ($D_1$--$D_3$) and the different depths of rectum canal (1 cm--5 cm) were revealed by the Sample Entropy. 

Definitely the most valuable information is provided by the analysis of $SampEn$ over multiple time scales. Through the 
interpretation of the mean MSE curves the stages of the most visible differences between AR and LAR groups were identified 
one month after operation $D_2$ for respectively 1 cm depth at the rest and 5 cm depth in the case of the MVC. That corresponds well to the clinical data as the LARS syndrome has its peak severity right after surgery with 
the diminishing frequency and severity months after the treatment \cite{walkega2001odlegle,wadja2002proceedings}. Also, since the amplitude of the sEMG signal depends 
on the distance between its source and the electrode and that distance is the smallest for superficial part of an 
external anal sphincter in resting conditions, the sEMG signal is almost always the strongest in the most external recordings. 
While in maximum contraction conditions, when the amplitude of the signal rises, the probability that the signal from deeper parts 
of the muscle will be sufficiently represented in the recording also rises. 

It is shown that the information carried by the sEMG signals measured one year after the surgery $D_3$ returns to the state of the 
first examination $D_1$ for the selected cases. This situation is also confirmed in a clinical practice since for those patients 
who improve, the return of a normal function happens within the period of the 1st year \cite{giandomenico2015quality, couwenberg2018effect} . 
Probability of later recovery is typically very small. 
The data acquired one month after the operation $D_2$ are also characterized by the lower values of $SampEn$ for the majority of cases
for the large time scales in the LAR group which can indicate to the greater impact of adverse phenomena associated with postoperative 
side effects in this very group of patients. 
Study of the stages before and one month after surgery in patients with AR and LAR individually show more visible differences for the 
latter group with the decrease of $SampEn$ values at the $D_2$ stage. 
Statistically significant differences are observed among almost all of the compared stages of treatment as well as the various 
rectum canal depths in both AR and LAR groups. Nonetheless, the group of patients who underwent the LAR is definitely more 
diversified based on the MSE. That may indicate different degree of injury to a neuromuscular system resulting from a 
multimodal treatment of those patients. Correlation of those changes with results of functional testing are lacking 
thus making further conclusions speculatory. 

The main limitations of this study are due to the problem of inter--subject variability. The large diversity in distribution of 
EAS innervation zones, mainly caused by the high level of the individual asymmetries, significantly affects the differences 
between the compared groups. This phenomenon is further strengthened by the diversity of signals within a single subject. 
The values of the entropy for the time series detected at one of three separated rings which consists of 16--channels each 
indicate relatively high variability over these channels. That discrepancy consists of many factors including the concept 
of \textit{weighted innervation zones}. Some of the innervation zones may have a greater importance than the others
because of the different sizes of motor units \cite{merletti2016surface}. We are not able to specify the series that characterize 
such dominant zones, therefore the results are averaged over all channels. Despite the relatively small values of standard 
deviations, an effect of inter--channels variability significantly influences the final results.

\vspace{6pt} 

\noindent {Authors would like to thank Andrew Watson for proofreading of this manuscript.}

\noindent {The following abbreviations are used in this manuscript:\\

\noindent 
\begin{tabular}{@{}ll}
EAS & external anal sphincter\\
sEMG & surface electromyography\\
AR & anterior resection\\
LAR & lower anterior resection\\
MVC & maximum voluntary contraction
\end{tabular}}


\end{document}